\documentclass[eps,PPCF,reprint,floatfix,superscriptaddress,showpacs,amsfonts
,amssymb,amsmath,preprintnumbers]{revtex4-1}
\usepackage{bm}
\usepackage{epsfig}
\usepackage{amsmath}
\usepackage{amssymb}
\usepackage{graphicx}
\usepackage{bm}
\usepackage{color}
\usepackage{subfigure}
\usepackage{lipsum}
\usepackage{appendix}

\newcommand{\fig}[1]{Fig.~\ref{#1}}

\newcommand{\eq}[1]{Eq.~(\ref{#1})}
\newcommand{\eqs}[2]{Eqs.~(\ref{#1}--\ref{#2})} 
\newcommand{\eqsc}[2]{Eqs.~(\ref{#1},~\ref{#2})} 
\newcommand{\Eq}[1]{Eq.~(\ref{#1})}

\newcommand{\be}{\begin{equation}}
\newcommand{\ee}{\end{equation}}

\newcommand{\bem}{\begin{multline}}
\newcommand{\eem}{\end{multline}}

\newcommand\bea{\begin{eqnarray}}
\newcommand\eea{\end{eqnarray}}

\newcommand{\curl}{\bm \nabla \times}

\begin{document}
\setcounter{secnumdepth}{1}

\title{General kinetic solution for the Biermann battery with an associated pressure anisotropy generation}

\author{K. M. Schoeffler}
\affiliation{GoLP/Instituto de Plasmas e Fus\~ao Nuclear,
Instituto Superior T\'ecnico,\\ Universidade de Lisboa, 1049-001 Lisboa, Portugal}
\author{L. O. Silva}
\affiliation{GoLP/Instituto de Plasmas e Fus\~ao Nuclear,
Instituto Superior T\'ecnico,\\ Universidade de Lisboa, 1049-001 Lisboa, Portugal}

\date{\today}

\begin{abstract}

Fully kinetic analytic calculations of an initially Maxwellian distribution with arbitrary density and temperature gradients exhibit the development of temperature anisotropies and magnetic field growth associated with the Biermann battery. The calculation, performed by taking a small order expansion of the ratio of the Debye length to the gradient scale, predicts anisotropies and magnetic fields as a function of space given an arbitrary temperature and density profile. These predictions are shown to qualitatively match the values measured from particle-in-cell simulations, where the development of the Weibel instability occurs at the same location and with a wavenumber aligned with the predicted temperature anisotropy.

\end{abstract}


\maketitle


\section{Introduction.}

Intense magnetic fields generated in laser-solid interaction laboratory
experiments~\cite{Stamper71,Li07,Gregori12,Gao15} as well as the seed field
required for the generation of astrophysical magnetic
fields~\cite{Kulsrud92,Kulsrud08} have been attributed to the Biermann
battery~\cite{Biermann50}. The Biermann battery mechanism generates magnetic
fields due to misaligned temperature and density gradients. Until recently, the
theory behind this mechanism has been restricted to fluid models where an extra
non-ideal term is added to Ohm's law.  Fluid models typically assume
significant collision rates that maintain the pressure tensor in the form of a
scalar during relevant time scales \footnote{In systems with large magnetic
fields (not relevant in this work) fluid models with a non-scalar pressure
tensor aligned with the field can be formulated.}.  These conditions are often
not present in astrophysical environments and thus a fully kinetic model is
necessary. 

The Biermann battery has been investigated with fully self-consistent kinetic
3D simulations~\cite{Schoeffler14, Schoeffler16}, and recently an analytical
model has been presented for the special case with initial density and
temperature gradients, which are perpendicular to each other
\cite{Schoeffler2017}.  In this previous paper, the kinetic equivalent of the
Biermann battery was demonstrated along with the purely kinetic effect of the
generation of a temperature anisotropy in the pressure tensor, where both
effects were shown to be relevant for a wide variety of settings including
astrophysical shocks and laser experiments with small collision rates. We will
hereupon refer to Ref.~\cite{Schoeffler2017} as Paper I.

Here, we present a more general model that describes, given an arbitrary
density and temperature profile, both the Biermann battery and the evolution of
the pressure tensor as a function of time and space. With this description, the
Biermann battery can be used to describe many weakly collisional scenarios.
Furthermore, the full evolution of the kinetic pressure tensor allows for a
measure of the magnitude and direction of the temperature anisotropies again as
a function of both time and space.  These anisotropies give rise to kinetic
instabilities such as the Weibel instability~\cite{Weibel59} seen
in~\cite{Schoeffler14, Schoeffler16} or instabilities that inhibit the heat
flux~\cite{Levinson92,Gary00}, and thus both this temperature anisotropy and the
kinetic Biermann battery are relevant for a wide variety of settings from
astrophysical shocks to laser experiments with small collision rates.  

\section{Model.}
In Paper I~\cite{Schoeffler2017} a few assumptions have been considered to
simplify the solution.  We show that  we can relax some of these assumptions to
make the solution general and applicable to many arbitrary systems.  First we
assumed that the temperature gradient was perpendicular to the density
gradient.  Second we solved for second order terms, but ignored second order
contributions to the initial temperature and density gradients.

Here the time evolution of the distribution function and
electromagnetic fields is solved from the coupled Vlasov
and Maxwell's equations.  We assume that only the electrons play a
role and the ions are static, only acting as a neutralizing background, and
begin with a instantaneously perturbed Maxwellian electron distribution:
\begin{equation}
\label{maxwellian}
f_M = n_0 {\left(\frac{1}{2\pi v_{T0}^2}\right)}^{3/2} 
\exp\left(-\frac{1}{2}\frac{v^2}{v_{T0}^2}\right) 
\text{,}
\end{equation}
where the instantaneous perturbation, like in Paper I~\cite{Schoeffler2017}, is
obtained by replacing $n_0$ and $v_{T0}$ with $n$ and $v_{T}$. However, rather
than choosing simple linear and perpendicular gradients, here we examine the
most general case, including up to second order gradients with arbitrary
angles:  
\be
n = n_0\left(1 + \epsilon x + \frac{1}{2}\epsilon^2\kappa_{nij} x_i x_j\right)
\text{,}
\ee
\be
v_{T} = v_{T0} \sqrt{1 + \delta_\parallel x +\delta_\perp y + \frac{1}{2}\delta^2\kappa_{Tij} x_i x_j}\text{,}
\ee
\be
\label{epsilondef}
\epsilon \equiv \frac{\lambda_D}{L_n} \equiv \frac{\lambda_D}{n}\frac{\partial n}{\partial x}\left(0\right)
\text{, }
\epsilon^2 \kappa_{nij} \equiv \frac{\lambda_D^2}{n}\frac{\partial^2 n}{\partial x_i\partial x_j}\left(0\right)
\text{,}
\ee
\be
\label{deltadef}
\delta_\parallel \equiv \frac{\lambda_D}{L_{T\parallel}} \equiv \frac{\lambda_D}{T}\frac{\partial T}{\partial x}\left(0\right)
\text{, }
\delta_\perp \equiv \frac{\lambda_D}{L_{T\perp}} \equiv \frac{\lambda_D}{T}\frac{\partial T}{\partial y}\left(0\right)
\text{,}
\ee
\be
\label{kappadef}
\delta^2 \kappa_{Tij} \equiv \frac{\lambda_D^2}{T}\frac{\partial^2 T}{\partial x_i\partial x_j}\left(0\right)
\text{.}
\ee
Note that $\delta$ defined by the gradient scale of the temperature has been
divided into two components $\delta^2 = \delta_\parallel^2 + \delta_\perp^2$,
corresponding to the gradients parallel and perpendicular to the density
gradient.

In this system, $\bm{v}$ is normalized to $v_{T0}$, $t$ to $\omega_{pe}^{-1}$,
and $\bm{x}$ to $\lambda_D$, where $\omega_{pe}$ is the plasma frequency for
density $n = n_0$, and $\lambda_D \equiv v_{T0}/\omega_{pe}$ is the Debye
length. We normalize the fields $E$ and $B$ to $E_0 \equiv m_e v_{T0}
\omega_{pe}/e$ and $B_0 \equiv m_e c \omega_{pe}/e$ respectively.  We assume
that $\epsilon$ and  $\delta$ are small and comparable to each other.
Furthermore, we take  $\epsilon^2 \kappa_{nij}$ and $\delta^2 \kappa_{Tij}$ to
be comparable to $\epsilon^2$.

Assuming $\bm x \sim \epsilon^0$, the initial distribution function to second order in $\epsilon$ and $\delta$ is:
\begin{multline}
	\label{f0}
	f_0 = f_M 
	+ \epsilon xf_M 
	- \frac{1}{2} \left(\delta_\parallel x  + \delta_\perp y\right) \left(3-v^2\right) f_M\\
	+ \frac{1}{2} \epsilon^2 \kappa_{nij} x_i x_j f_M
	-  \frac{1}{4} \delta^2 \kappa_{Tij} x_i x_j \left(3- v ^2\right) f_M\\ 
	+ \frac{1}{8} {\left(\delta_\parallel x + \delta_\perp y \right)}^2\left(15 -10 v ^2 + v ^4 \right) f_M\\
	-  \frac{1}{2} \epsilon x  \left(\delta_\parallel  x + \delta_\perp  y \right) \left(3- v ^2\right) f_M 
	\text{.}
\end{multline}
We evolve the Vlasov-Maxwell equations initialized with this distribution
function, and either no initial electric or magnetic fields, or natural
equilibrium fields that balance the pressure gradient.  The steady state
electric field for the complete general solution is the following.
\begin{align}
\label{Esteady}
\bm{E} = \bm{E_{st}} \equiv
 -
\left(\epsilon
+\delta_{\parallel} 
-\epsilon^2
x 
+
\kappa_{nix}\epsilon^2
x_i 
+
\kappa_{Tix}\delta^2
x_i 
 \right.
\nonumber \\
\left.+\epsilon\delta_{\parallel}
x 
+ \epsilon\delta_{\perp}
y 
 \right)
 \hat{\bm x}&
\nonumber \\
-
\left(
\delta_{\perp}
+ \kappa_{niy}\epsilon^2
x_i
+ \kappa_{Tiy}\delta^2
x_i 
\right)
\bm{\hat{y}}&
\nonumber \\
-
\left(
\kappa_{niz}\epsilon^2
x_i
+ \kappa_{Tiz}\delta^2
x_i 
\right)
\bm{\hat{z}}&
\text{,}
\end{align}

Note that the perturbation in \eq{f0} is taken as a given initial state.  The
Biermann battery is not an instability (in fluid models it grows linearly with
time; we show here this remains true in the kinetic case), and therefore it only occurs with non-equilibrium conditions.

The total distribution function evolves as:
\begin{equation}
\label{vlasov}
\frac{\partial f}{\partial t} + \mathbf{v} \cdot \bm \nabla f 
- \left( \mathbf{E}
+ \mathbf{v}Â \times \mathbf{B} \right)
\cdot \bm \nabla_{v} f = 0
\text{,}
\end{equation}
\begin{equation}
\label{faraday}
\frac{\partial \mathbf{B}}{\partial t} = -\curl \mathbf{E}
\text{,}
\end{equation}
\begin{equation}
\label{ampere}
\frac{\partial \mathbf{E}}{\partial t} = \int dv^3 \mathbf{
v} f + \frac{c^2}{v_{T0}^2} \curl \mathbf{B}
\text{,}
\end{equation}
where $\mathbf{\nabla_{ v }}$ is the gradient in velocity space,
\eq{vlasov} is the Vlasov equation, \eq{faraday} is Faraday's law, and
\eq{ampere} is Ampere's law. 

Following the same assumptions made in Paper I~\cite{Schoeffler2017}, we seek
solutions to these equations in powers of $\epsilon$ and $\delta$.  We assume
$t \sim  \bm{x} \sim c^2/v_{T0}^2 \sim \epsilon^0 \sim \delta^0$.  Although the
solution is only valid when $\bm{x} \sim \epsilon^0$, at an arbitrary position
$\bm{x}$, the calculation also remains valid with a normalization based on the
local $v_{T}$ and $n$. In some regions where the local $\epsilon$ and  $\delta$
go to zero, it is the $\epsilon^2 \kappa_{nij}$ and $\delta^2 \kappa_{Tij}$
that remain as the small parameters.  Besides $\epsilon$ and $\delta$, three
other parameters $c_s/v_{T0}$, $v_{T0}^2/c^2$, and $\nu/\omega_{pe}$, where
$c_s$ is the sound speed, and $\nu$ is the collision frequency must remain
small. Each of these parameters are assumed to be much smaller than one, but
aside from $\nu/\omega_{pe}$ can in principle remain of order $\epsilon^0$.  In
fact, our calculation assumes $v_{T0}^2/c^2 \sim \epsilon^0$, although as long
as $\bm B \sim \epsilon^2$ (which we find in our solution), it is acceptable
for $v_{T0}^2/c^2 \sim \epsilon^1$. Small values for these parameters are
implicitly assumed when considering static ions, using the non-relativistic
Vlasov equation/ Maxwellian distribution, and neglecting collisions.

As explained in Paper I~\cite{Schoeffler2017}, for the first order solution no
magnetic field is generated, and only bulk flows and temperature fluxes could
be obtained from the distribution function. It was necessary that we perform
our calculation with second order terms ($\sim \epsilon^2$) to see effects
including the Biermann battery, and the formation of a temperature anisotropy.
Once again it should emphasized that modifications coming from $c_s/v_{T0}$ and
$v_{T0}^2/c^2$ can be neglected for both first order and second order
solutions, although the terms from the first order solution are then only
accurate to $\epsilon^1$.

\section{Density gradient.}
For the generalized scenario studied here, we first consider the case with only
a density gradient ($\delta=0$). If we assume the initial condition of $f =
f_0$ and no initial electric or magnetic fields, a solution can be found taking
an expansion for small $t$, restricted to second order in
$\epsilon$.  At this point the only difference from the calculations done in
Paper I~\cite{Schoeffler2017} is that we include the second order gradients
$\kappa_{nij}$.

Following Paper I~\cite{Schoeffler2017}, summing over all orders of $t$ converges to the analytic
solution valid for $t \sim \epsilon^0$:
\be
f = f_0 + \tilde{f_n}
\ee
\begin{equation}
\label{perturbedEdensity}
\bm{E} = \bm{E_{st}}
\left[1-\cos\left(\omega_{pe,x}t\right)\right]
\text{,}
\end{equation}
where,
\begin{multline}
\label{perturbedfdensity}
\tilde{f_n} \equiv
- \epsilon  \sin\left(\omega_{pe,x}t\right) v_x  f_M \\
- \epsilon^2 \kappa_{nij}  \sin\left(\omega_{pe,x}t \right) x_j v_i  f_M\\
+\frac{1}{2}
\epsilon^2
 \sin\left(\omega_{pe,x}t\right) 
x  
 v_x  
 f_M\\
+\epsilon^2 
\left[1-\cos\left(\omega_{pe,x}t\right)
\right]
\left(\kappa_{nij}  v_i   v_j  - v_x^{2}\right) 
f_M\\
+\frac{1}{2}\epsilon^2
  t \sin\left(\omega_{pe,x}t\right)
  v_x^{2}
f_M\\
-\frac{1}{2}\epsilon^2 
{\left[1-\cos\left(\omega_{pe,x}t \right)\right]}^2
\left( v_x^{2} -1\right) 
f_M
\text{,}
\end{multline}
and  $\omega_{pe,x}=1+\epsilon x/2$ is the
plasma frequency based on the $x$ dependent density, $n$.  Note that we have
made use of the Poincar\'e-Lindstedt method~\cite{Lindstedt1882,Poincare1893},
which by modifying the frequency in the solution, avoids unphysical secularly
growing terms. This is done by including additional higher order terms ($>
\epsilon^2$) found in the expansion of the sin and cos terms. It is evident
that the electric field of this solution oscillates about $\bm{E_{st}}$.

Despite the use of the Poincar\'e-Lindstedt technique, there still exists a secular term
in $f$, which grows linearly with time, and eventually grows beyond
$t=\epsilon^{-1}$, and breaks the assumptions
of the ordering. Thus our model is only valid as long as $t$ remains small
compared to this limit. This term is, however, physical, and represents the
increasing electron density associated with the divergence of the electric
field: 
\begin{multline}
\bm \nabla \cdot \mathbf{E} = (n_i-n_e)/n_0\\ =
\epsilon^2 
 \left[1 - \cos\left(\omega_{pe,x}t\right) - \frac{1}{2}\omega_{pe,x}t \sin\left(\omega_{pe,x}t\right) \right]
\text{.}
\end{multline}
The space dependent frequency, $\omega_{pe,x}$, gives rise to increasingly
shorter scale variations along $x$ in the electric field, and thus an
increasingly large divergence. These variations along $x$ lead to phase mixing
in space and then Landau damping. This damping at early times is exponentially
repressed, and does not show up in our expansion. However, when the damping
becomes most significant at $k\lambda_D \sim 1$ (equivalent to $t
\sim \epsilon^{-1}$) the assumptions break down. Eventually Landau damping
eliminates both the oscillations and the secular term, and thus the electric
field should naturally settle to \eq{Esteady}. If we take \eq{Esteady} as the
initial condition for the electric field, we arrive at a simple equilibrium
solution to \eqs{vlasov}{ampere} where $\bm{E}$ and $f$ do not change with
time. 

\section{Temperature gradient.}
We now consider a second case, with only a temperature gradient
($\epsilon =0$, $\delta\ne0$).  
For simplicity, at this point
we will move to a reference frame where $\delta_\parallel = \delta$, and $\delta_\perp=0$.
In this reference frame aligned with the temperature gradient,
\be
\label{xprime}
\delta x^\prime = \delta_\parallel x  + \delta_\perp y
\ee
and
\be
\label{vxprime}
\delta  v_x^\prime = \delta_\parallel  v_x   + \delta_\perp  v_y.
\ee
Note that when looking at only a temperature gradient we are free to use this
reference frame; however, when combining both gradients it is important to
return to the original coordinates, replacing $x^\prime$ and $v_x^\prime$ with
\eqs{xprime}{vxprime}.

If we again start with the initial conditions,
$f = f_0$, and no initial electric or magnetic fields, the solution to \eqs{vlasov}{ampere} is, to second order in
$\delta$, the following:
\be
f = f_{\nabla T} + \tilde{f_T} 
\ee
\be
\label{perturbedEpressure}
\bm{E} = \bm{E_{st}}
\left[1-\cos\left(\omega_{pe,x}t\right)\right]
\text{,}
\ee
where
\begin{multline}
\label{fTevolution}
f_{\nabla T} \equiv f_0 
+ \frac{1}{2}\delta \omega_{pe,x}t v_x^\prime \left(5 - v ^2\right) f_M\\
-\frac{1}{4}\delta^2 \omega_{pe,x}t x^\prime v_x^\prime  \left(25 - 12 v ^2 +  v ^4 \right) f_M\\
+\frac{1}{2} \delta^2 \kappa_{Tij} \omega_{pe,x}t x_i  v_j  \left(5 - v ^2\right) f_M\\
+ \frac{1}{8} \delta^2  
{(\omega_{pe,x}t)}^2 {v_x}^{\prime2} \left(25 - 12 v^2 +  v ^4\right) f_M\\
+ \frac{1}{4} \delta^2 {(\omega_{pe,x}t)}^2 \left[ v_x^{\prime2} \left(7 -  v ^2\right) -  \left(5 - v ^2\right)\right] f_M\\
 - \frac{1}{4} \delta^2  \kappa_{Tij} {(\omega_{pe,x}t)}^2  v_i   v_j \left(5 - v ^2\right) f_M
\text{,}
\end{multline}
and
\begin{multline}
\tilde{f_T} \equiv
- \delta \sin\left(\omega_{pe,x}t \right) v_x^\prime f_M\\
+\frac{1}{2}\delta ^2 \sin\left(\omega_{pe,x}t \right) x^\prime  v_x^\prime \left(5- v ^2\right) f_M\\
- \delta^2 \kappa_{Tij} \sin\left(\omega_{pe,x}t \right)  x_j v_i  f_M\\
-\frac{1}{2} \delta^2  \left[1-\cos\left(\omega_{pe,x}t \right)\right]v_x^{\prime2} \left(5- v ^2\right)f_M\\
+\frac{1}{2}\delta^2\left[1-\cos\left(\omega_{pe,x}t \right)\right]\left[ v_x^{\prime2} \left(7- v ^2\right) -\left(5 -  v ^2\right)\right] f_M \\
-\frac{1}{2} \delta^2 \omega_{pe,x}t \sin\left(\omega_{pe,x}t \right)\left[ v_x^{\prime2} \left(7- v ^2\right) - \left(5 -  v ^2\right)\right]f_M \\
+\frac{1}{2} \delta^2 {\left[1-\cos\left(\omega_{pe,x}t \right)\right]}^2 \left(1 -  v_x^{\prime2} \right)f_M\\
+\delta^2 \kappa_{Tij} \left[1-\cos\left(\omega_{pe,x}t \right)\right]  v_i   v_j  f_M \\
\text{,}
\end{multline}

Although in principle there is nothing to damp these oscillations, if it were a
system with density gradients it would Landau damp as described earlier, and in
Paper I~\cite{Schoeffler2017}.  As before, we consider \eq{Esteady} as the initial
condition.  This yields a simpler solution where the electric field is constant
with time, but it still allows the distribution function to evolve with time as
$f = f_{\nabla T}$. 

Like the scenario with a density gradient, there are terms proportional to $t$ in the distribution
function $f_{\nabla T}$, which eventually break the assumptions of the
ordering. The second term on the RHS of \eq{fTevolution} is associated with the
heat flux, and matches the collisional solution shown in~\cite{Levinson92} once
$t$ reaches the collision time, unless this term breaks the assumptions of the
ordering first, once $\omega_{pe}~=\delta^{-1}$.  If we integrate over $1$,
$v_i$, and $v_i v_j$, we find the density remains $n$, and the flow remains
$0$, but the temperature tensor changes.

In Paper I~\cite{Schoeffler2017}, an anisotropy in the temperature tensor which
was hotter in the direction of the temperature gradient
($T_{xx}^\prime>T_{yy}^\prime$) was shown to grow proportional to $\delta^2
t^2$, an so the temperature gradient naturally lead to a temperature
anisotropy, which gives rise to kinetic instabilities such as the Weibel
instability~\cite{Weibel59} seen in~\cite{Schoeffler14} or can drive instabilities that
inhibit the heat flux~\cite{Levinson92,Gary00}.
Again we define the temperature tensor as:
\be
\label{temperaturedef}
\frac{T_{ij}}{m_e v_{T0}^2} \equiv \frac{1}{n} \int dv^3  v_i  v_j  f \text{,} 
\ee
The third to last term on the RHS of \eq{fTevolution}, which grows as $t^2$, is
associated with this temperature anisotropy.
The temperature tensor (in the temperature gradient aligned frame) is obtained by integrating
\eq{fTevolution} over $v_i v_j$:
\be
T_{ij} = v_T^2 \mathbb{I}  + T_{\nabla T, ij},
\ee
where the change in the temperature due to the temperature gradient:
\be
\label{deltapresstens}
T_{\nabla T, ij}^\prime \equiv 
\frac{1}{2}\delta^2\left(\omega_{pe,x}t\right)^2
\left[
 \begin{pmatrix}
	 3& 0& 0\\
	 0& 1& 0\\
	 0& 0& 1 
 \end{pmatrix} 
 +2 \kappa_{Tij} + \text{Tr}\left(\kappa_{Tij}\right) \mathbb{I}
 \right]
\ee
Note that the primed notation indicates that this tensor is presented in the
temperature gradient aligned frame.  Before doing further analysis on this
anisotropic pressure tensor, we will also look into the effects of both
$\epsilon$ and $\delta$.

\section{Biermann battery.}
We now consider a third case, with both gradients
($\epsilon \ne 0$, $\delta\ne0$).  If we again start with the initial conditions,
$f = f_0$, and no initial electric or magnetic fields, to second order in
$\delta$ and $\epsilon$, the solution to \eqs{vlasov}{ampere} is the following:
\be	
\label{biermannfosc}
	f = f_{\nabla n, \nabla T}  + \tilde{f_n} + \tilde{f_T} + \tilde{f}_{n,T} \text{,}
\ee
\be
\label{biermannEosc}
\bm{E} = \bm{E_{st}}
\left[1-\cos\left(\omega_{pe,x}t\right)\right]
\ee
\be
\bm{B}
\label{biermannBosc}
= -
\epsilon\delta_\perp
\left[\omega_{pe,x}t -\sin\left(\omega_{pe,x}t \right) \right]
\bm{\hat{z}}
\text{.}
\ee
where,
\begin{multline}
	\label{fbiermann}
	f_{\nabla n, \nabla T} \equiv
	f_{\nabla T} +  
	\frac{1}{2}\epsilon  
	\left(\delta_\parallel   v_x  + \delta_\perp  v_y  \right)
	\omega_{pe,x}t 
	 x \left(5 - v ^2\right) f_M\\
	 - \frac{1}{4}\epsilon\delta_\parallel 
	 {\left(\omega_{pe,x}t\right)}^2 
	 \left(3 -  v ^2\right) f_M\\
	 - \frac{1}{2}\epsilon\delta_\parallel
	 {\left(\omega_{pe,x}t\right)}^2 
	 \left(1- v_x ^2\right) f_M\\
	 + \frac{1}{2}\epsilon\delta_\perp  
	 {\left(\omega_{pe,x}t\right)}^2 
	  v_x   v_y  f_M
	  \text{,}
\end{multline}
and
\begin{multline}
\label{perturbedfpressure}
\tilde{f}_{n,T} \equiv\\
-\epsilon  \left(\delta_\parallel x+ \delta_\perp y\right)  \sin\left(\omega_{pe,x}t \right) v_x  f_M\\
-\frac{1}{2}\epsilon \left(\delta_\parallel v_x  + \delta_\perp v_y   \right)  \sin\left(\omega_{pe,x}t \right) x f_M\\
+\frac{1}{2}\epsilon  \left(\delta_\parallel x+ \delta_\perp y\right)  \sin\left(\omega_{pe,x}t \right) v_x \left(5- v ^2\right) f_M\\
+2\epsilon \left(\delta_\parallel  v_x   + \delta_\perp  v_y \right) \left[1-\cos\left(\omega_{pe,x}t \right)\right] v_x f_M\\
-\frac{1}{2}\epsilon \delta_\parallel \left[1-\cos\left(\omega_{pe,x}t \right)\right]\left(5- v ^2\right)f_M\\
+ \epsilon  \left(\delta_\parallel v_x  + \delta_\perp v_y \right) {\left[1-\cos\left(\omega_{pe,x}t \right)\right]}^2
 v_x  f_M \\
- \epsilon \delta_\parallel {\left[1-\cos\left(\omega_{pe,x}t \right)\right]}^2f_M \\
-\frac{1}{2}\epsilon \left(\delta_\parallel v_x  + \delta_\perp  v_y  \right)\omega_{pe,x}t \sin\left(\omega_{pe,x}t \right) v_x \left(7- v ^2\right)f_M \\
+\frac{1}{2}\epsilon\delta_\parallel  \omega_{pe,x}t \sin\left(\omega_{pe,x}t \right) v_x^2 f_M \\
+\frac{1}{2}\epsilon\delta_\parallel \omega_{pe,x}t \sin\left(\omega_{pe,x}t \right)\left(5- v ^2\right)f_M
\text{,}
\end{multline}

Once again the solution oscillates at $\omega_{pe}$ time scales 
around a steady state. 

Starting with the steady state electric fields, \eq{Esteady}, we arrive at: 
\be
\label{biermannf}
f = f_{\nabla n, \nabla T} \text{,}
\ee
\be
\label{biermannE}
\bm{E} = \bm{E_{st}}
\ee
\be
\label{biermannB}
\bm{B}
= -
\epsilon\delta_\perp \omega_{pe,x}t 
\bm{\hat{z}}
\text{.}
\ee

We thus see a kinetic solution of the growth of magnetic fields via the
Biermann battery, which grows linearly with time and proportional to the cross
product between the density and temperature gradients (see \eq{biermannB}), confirming the fluid model
prediction.

The third and fourth terms of~\eq{fbiermann}, modify the isotropic temperature (Tr$\left(T_{ij}\right)$),
and the temperature in the $\hat{x}$ direction ($T_{xx}$), respectively.  The last term is
associated with the off-diagonal component of the pressure tensor
($T_{xy}$), which enhances and rotates the magnitude and direction of the
temperature anisotropy (defined in the frame that diagonalizes $T_{ij}$).

If we integrate \eq{biermannf} over $v_i v_j$, we find how the pressure tensor is modified:

\be
\label{addpresstens}
T_{\nabla n \nabla T,ij} = 
\frac{1}{2}\epsilon\left(\omega_{pe,x}t\right)^2
 \begin{pmatrix}
	 \delta_\parallel& \delta_\perp     & 0 \\
	 \delta_\perp&     3\delta_\parallel& 0 \\
	 0&                0&               \delta_\parallel
 \end{pmatrix} 
\text{.}
\ee
If we rotate this into the temperature gradient aligned frame (again indicated by the primed notation), in order to
compare with \eq{deltapresstens}, it makes more sense to employ
$\epsilon_\parallel = \epsilon \delta_\parallel/\delta$, and $\epsilon_\perp
= -\epsilon \delta_\perp/\delta$, such that
\be
\label{addpresstensTF}
T_{\nabla n \nabla T,ij}^\prime = 
\frac{1}{2}\delta\left(\omega_{pe,x}t\right)^2
 \begin{pmatrix}
	 3\epsilon_\parallel& \epsilon_\perp     & 0 \\
	 \epsilon_\perp&      \epsilon_\parallel & 0 \\
	 0&                   0                  & \epsilon_\parallel
 \end{pmatrix} 
\text{.}
\ee

The total pressure is now given by:
\be
\label{totpresstens}
T_{ij} =  v_T^2 \mathbb{I} +
T_{\nabla T, ij} + T_{\nabla n \nabla T,ij} 
\text{.}
\ee

If we diagonalize the in-plane components (plane including $\bm \nabla n$ and $\bm \nabla T$) of the matrix from \eq{totpresstens}, the diagonal terms are:
\begin{align}
	\label{doubleprime}
	T_{xx}^{\prime\prime} =&  v_T^2 + \Delta T_0 + \frac{1}{2}A_0\text{,}\nonumber\\ 
	T_{yy}^{\prime\prime} =&  v_T^2 + \Delta T_0 - \frac{1}{2}A_0\text{,}\nonumber\\ 
	T_{zz}^{\prime\prime} =&  v_T^2 + \Delta T_0 - \frac{1}{2}\Delta T_0 \text{,}
\end{align}
where
$T_{xx}^{\prime\prime}$ is the
hotter direction, $T_{yy}^{\prime\prime}$ is the cooler direction, $\Delta T_0$ is the
average increase of these two in-plane components of the temperature, and $A_0$ is the
in-plane temperature anisotropy. For a 2D system where the derivatives in the $\hat{z}$ direction
are zero the in-plane temperature and anisotropy can be expressed as:
\be
\Delta T_0= \left(\delta^2 + \delta^2 \left(\kappa_{Txx}^\prime + \kappa_{Tyy}^\prime\right) + \epsilon_\parallel  \delta\right) \left(\omega_{pe,x}t\right)^2 \\
\ee
\begin{multline}
\label{genanisotdef}
	A_0 = \left[\left(\delta^2 +\delta^2 \left(\kappa_{Txx}^\prime -  \kappa_{Tyy}^\prime\right)  + \epsilon_\parallel  \delta\right)^2 \right. \\
	     \left. + \left(2 \delta^2\kappa_{Txy}^\prime + \epsilon_\perp \delta\right)^2\right]^{1/2} \left(\omega_{pe,x}t\right)^2
\end{multline}
As long as $\Delta T_0 < A_0$, $A_0$ is the dominant anisotropy.  
One can observe that this anisotropy is proportional to the sum of 3 vectors,
\be
\label{vectoranisotdef}
A_0 = \delta |\bm{\delta} + \bm{\epsilon} + \delta \mathbf{\Delta}|\left(\omega_{pe,x}t\right)^2,
\ee
where
$\bm{\epsilon} = \bm \nabla n /n$,  $\bm{\delta} = \bm \nabla T/ T$, and
$\mathbf{\Delta}$ is a vector with magnitude $\Delta =
\left(\left(\kappa_{Txx}^\prime - \kappa_{Tyy}^\prime\right)^2 +
4\kappa_{Txy}^{\prime2}\right)^{1/2}$, where the components parallel and
perpendicular to the temperature gradient are $\Delta_\parallel =
\left(\kappa_{Txx}^\prime -\kappa_{Tyy}^\prime\right)$, and $\Delta_\perp = 2
\kappa_{Txy}^\prime$. The angle with respect to the temperature gradient, at which the temperature is the largest, can be expressed as:
\be
\label{angledef}
\theta^\prime = \frac{1}{2} \text{tan}^{-1}\left(\frac{\epsilon_\perp  + \delta \Delta_\perp}{\delta + \delta \Delta_\parallel + \epsilon_\parallel}\right)
+ \frac{\pi}{2}\left(1-\Theta\left(\delta + \delta \Delta_\parallel + \epsilon_\parallel\right)\right)
\ee
where $\Theta(x)$ is the step function.  Although the first term in
\eq{angledef} is restricted to  $-\pi/4 < \theta^\prime < \pi/4$, the direction
of the hotter temperature shifts by $\pi/2$ when the denominator changes sign
(i. e. the temperature aligned with the temperature gradient is cooler).  Note
that the temperatures parallel or anti-parallel to the temperature gradient are
equivalent, so it is justified that $\theta^\prime$ remains restricted between
$-\pi/4 < \theta^\prime < 3\pi/4$. We thus have all the information required to
take an arbitrary initial distribution of temperature and density gradients,
and solve for the time evolution of the temperature, anisotropy, and angle of
anisotropy.

Note that the solution is also valid for an arbitrary uniform magnetic field as
long as $\bm \epsilon$, $\bm \delta$, and $\bm \Delta$ are all parallel to the
magnetic field. The generation of this anisotropy will thus also occur in
magnetized plasmas, where different instabilities, such as the firehose
instability~\cite{Parker58}, would likely form. 

\section{Numerical Simulations.}
\begin{figure}
  \noindent\includegraphics[width=3.5in]{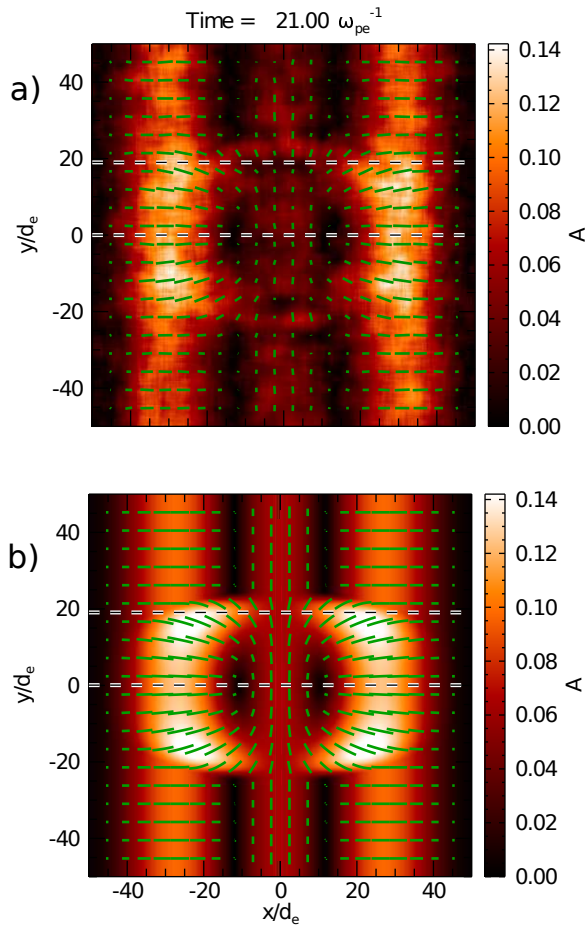}
  \caption{\label{Anisotropy}
	  Temperature anisotropy measured from PIC simulation (a) with $L_T/d_e=50$
  $(L_T/\lambda_D= 250)$ and $m_i/m_e=2000$, at $\omega_{pe}t=21$, reported in
  Ref.~\cite{Schoeffler16}. The green lines represent the direction ($i$) where
  the temperature ($T_{ii}$) is maximized.  The predicted anisotropy (b)
  given by \eqs{genanisotdef}{angledef} for the initial density and
  temperature distribution from \eq{distributionfunction} show qualitatively similar results.  The
  dashed lines represent the location of the line-outs presented in
  \fig{Acut}.
}
\end{figure}
\begin{figure}
  \noindent\includegraphics[width=2.5in]{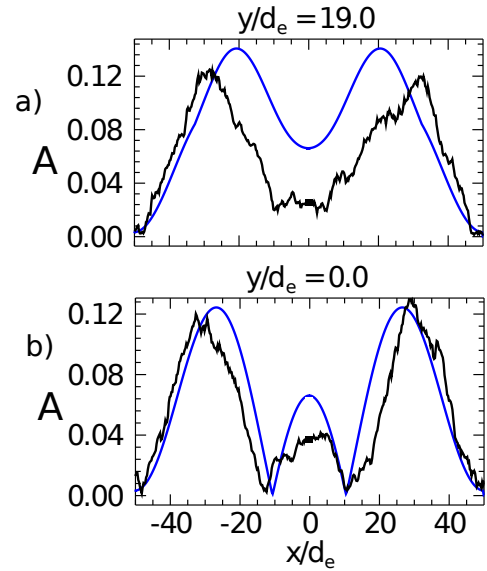}
  \caption{\label{Acut}
  Line-outs of the plots of the anisotropy, A, shown in \fig{Anisotropy} at
  $y/d_e =$ (a) 19, and (b) 0. The black curve is the simulation results
  (\fig{Anisotropy}(a)), the blue is the theoretical predictions assuming the
  initial density and temperature distribution (\fig{Anisotropy}(b)).
}
\end{figure}

A prediction of the anisotropy and Biermann fields generated at early times ($t
\ll \delta^{-1}$), can be obtained via \eqs{genanisotdef}{angledef} and
\eq{biermannB}.  The prediction can be compared directly with
particle-in-cell (PIC) simulations using the OSIRIS framework \cite{Fonseca02,
Fonseca08}, with the aid of a new diagnostic of the temperature tensor. 
We will thus compare two PIC simulations with this prediction;
one with $L_T/d_e=50$ $(L_T/\lambda_D= 250)$ and $m_i/m_e=2000$, reported in
Ref.~\cite{Schoeffler16}, and one based on simulations from
Ref.~\cite{Schoeffler14} ($L_T/d_e=200$ $(L_T/\lambda_D= 1000)$ and $m_i/m_e=25$),
but where we isolate the magnetic fields due to the Weibel instability by choosing
parallel density and temperature gradients both in the radial direction. 

In \fig{Anisotropy}(a), the anisotropy for the simulation with $L_T/d_e = 50$,
at $\omega_{pe}t = 21$ is shown. The anisotropy is calculated from the
temperature tensor as shown in Appendix~\ref{arbitraryanisot}.  For this
simulation, $\delta^{-1} \approx 250$ based on $L_T$ and the central density
and temperature ($v_{T0}/c = 0.2$). Since $\omega_{pe}t \ll \delta^{-1}$, in
principle our assumptions are not largely broken.  We solve for the small
parameters defined in \eqs{epsilondef}{kappadef} as a function of space using
the initial density and temperature distribution from the simulation (shown in
Appendix~\ref{nTdistribution}). Note that \eqs{epsilondef}{kappadef} assume a
frame aligned with the temperature gradient. In Appendix~\ref{arbitraryframe},
the transformations from an arbitrary frame are shown for reference.  

We find that the minimum $\delta^{-1} \approx 26$, $\epsilon^{-1} \approx 11$,
and $|\delta^2\kappa_{ij}|^{-1/2} \approx 26$, so in certain regions our
assumptions are only marginally held. For example in the most extreme case, by
$\omega_{pe}t=21$ the local value of $\omega_{pe}t\epsilon \approx 0.75$ (at
$x/d_e \approx 0$, $|y|/d_e \approx 20$).

\fig{Anisotropy}(b) shows the theoretically predicted spatial distribution of
anisotropy. Despite the marginal assumptions, the distribution matches
qualitatively quite well with the simulation results, with only slight
differences. The anisotropy around $|x|/d_e = 30$ expands outward slightly,
while the anisotropy growth around $x/d_e = 0$ is a bit suppressed during this
expansion. 

In order to get a more quantitative comparison, two cuts of the anisotropy at
$y/d_e=0$ and $19$ are shown in \fig{Acut}. The black curve is the simulation,
while the blue curve is the prediction for the initial simulation temperature
and density profile given in \eq{distributionfunction}. The predictions hold
very well, with only a slight departure near the region in the center, and a
remarkable quantitative agreement on the anisotropy.
\begin{figure}
  \noindent\includegraphics[width=3.5in]{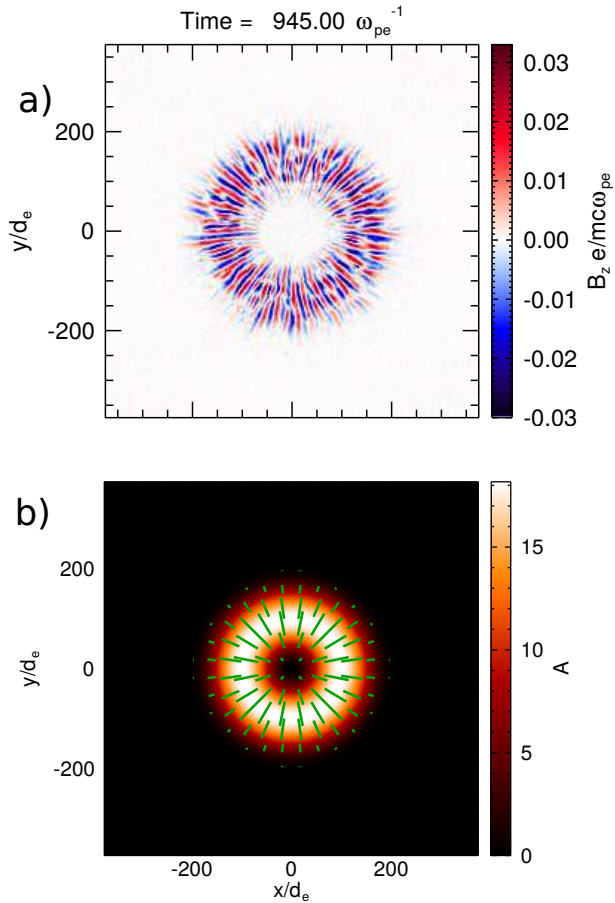}
  \caption{\label{Anisotropycircle}
  Out-of-plane magnetic field $B_z$ generated by the Weibel instability from the PIC
  simulation (a) based on simulations reported in Ref.~\cite{Schoeffler14}
  ($L_T/d_e = 200$ ($L_T/\lambda_D = 1000$), $m_i/m_e= 25$, at
  $t=945~\omega_{pe}^{-1}$), isolating the Weibel instability with radially
  parallel temperature and density gradients. The predicted temperature
  anisotropy and angle (b) given by \eqs{genanisotdef}{angledef} for the initial density and temperature 
  distribution from \eq{distributionfunction} coincides with the region
  where the Weibel instability occurs. The green lines represent the direction
  ($i$) where the temperature ($T_{ii}$) is maximized.
}
\end{figure}

For the simulation where we isolate the Weibel instability (based on simulations
from Ref.~\cite{Schoeffler14} with $L_T/d_e=200$ $(L_T/\lambda_D= 1000)$ and
$m_i/m_e=25$), by choosing parallel density and temperature gradients, $L_n =
L_T$ and the the temperature proportional to $r = \sqrt{x^2+y^2}$ instead of
$x$. The anisotropy is predicted to point radially outward, and thus only an
out-of-plane Weibel magnetic field ($B_z$) should grow with a wavenumber $k$ in
the azimuthal direction. In \fig{Anisotropycircle}(a), the magnetic fields
generated by the Weibel instability are shown.  Again we show the predicted
anisotropy for the initial density and temperature distribution from
\eq{distributionfunction}. As expected, the Weibel instability coincides with these
predicted anisotropy distributions that drive it. Note that the Weibel
instability occurs at a slightly larger radius as the anisotropy expands with
time.

\begin{figure}
  \noindent\includegraphics[width=3.5in]{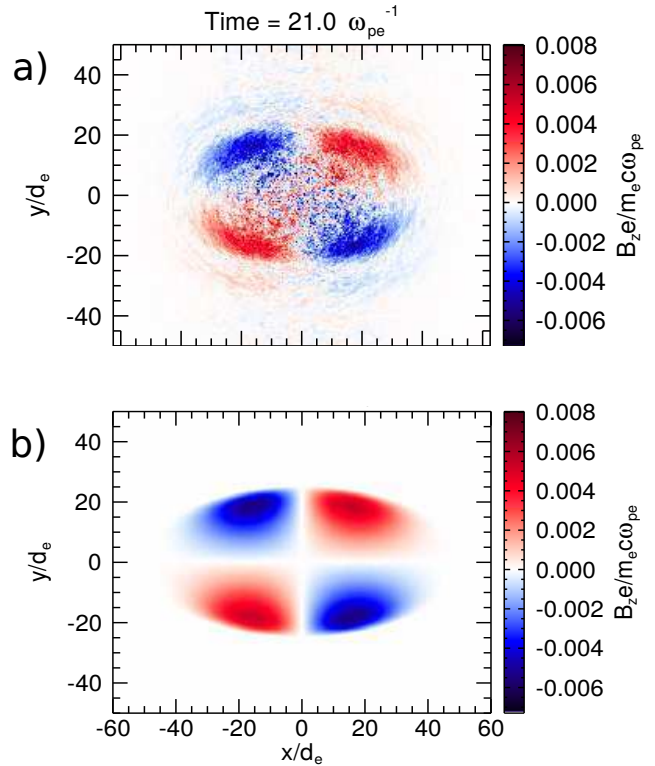}
  \caption{\label{Biermann}
	  Out-of-plane magnetic field $B_z$ generated by the Biermann battery from the 
	  PIC simulation (a) shown in \fig{Anisotropy} originally reported in Ref.~\cite{Schoeffler16}
	  ($L_T/d_e = 50$ ($L_T/\lambda_D = 250$), $m_i/m_e= 2000$, at $\omega_{pe}t=21$). The
	  predicted magnetic field (b) given by \eq{biermannB} for the
	  initial density and temperature distribution from \eq{distributionfunction} shows
	  agreement with the simulation results. 
  }
\end{figure}

Finally, in \fig{Biermann}(a), the magnetic field is shown from the simulation
also shown in \fig{Anisotropy}. This out-of-plane magnetic field ($B_z$) is
compared with the respective fields predicted in \eq{biermannB}, for the
initial density and temperature distribution from \eq{distributionfunction}.
Here, like in \fig{Anisotropy} we are looking at $\omega_{pe}t=21$, which
is much less than the crossing time $\omega_{pe}t\sim \delta^{-1} = 250$,
so our assumptions are not largely broken. Although the linear predictions are
not strictly accurate, again we find a good match with the fields from simulation.

\section{Conclusions.}
We have performed kinetic calculations with density and temperature gradients,
which predict the linear growth of magnetic fields collisional regimes, and the
generation of a temperature anisotropy driven by the temperature gradient,
which goes beyond the results from Paper I~\cite{Schoeffler2017} by allowing a
completely general set of gradients. With this generality, a spatial dependent
prediction of the temperature anisotropy (magnitude and direction) which
coincides with the Weibel instabilities that this anisotropy drives, and the
magnetic fields driven by the kinetic Biermann battery is now possible.

Similarly to the results reported in Paper I~\cite{Schoeffler2017}, the kinetic
outcome of the anisotropy generation is relevant even for some magnetized
cases; as long as there are no gradients perpendicular to $\bm B$, where the
magnetic field would affect the relevant particle motions.  This phenomena is
thus relevant for a wide variety of settings.

Likewise, for the more general case taken in this paper, the evolution of an
anisotropic Maxwellian distribution ($v_{Ti0} \ne v_{Tj0}$, where $v_{Ti0}$ is
the thermal velocity in the $i$ direction) can be modeled by the presented
equations.  In that case, $\bm x$, $\bm v$, and $\bm E$ are normalized using
the $v_{Ti0}$ in the same direction, and \eq{biermannB} has an additional
factor of $v_{Tx0}/v_{Ty0}$. This reduces to the Biermann field being generated due to the thermal velocity solely in the direction of the density gradient.

\section{Acknowledgments.}
This work was supported by the European Research Council (ERC-2010-AdG Grant
No. 267841, and ERC-2015-AdG Grant No. 695008). 

{\section{\appendixname}}
\begin{appendices}
\section{Arbitrary frame conversion}
	\label{arbitraryframe}
In order to calculate the anisotropy from \eq{genanisotdef} in an arbitrary
frame the following transformations to the temperature aligned frame are
provided below.
\be
\kappa_{xx}^\prime = \frac{\kappa_{xx} + \kappa_{yy}}{2}+\left(\kappa_{xx} - \kappa_{yy}\right) \frac{\delta_x^2-\delta_y^2}{2\delta^2} + \kappa_{xy} \frac{2\delta_x\delta_y}{\delta^2} 
\ee
\be
\kappa_{xx}^\prime = \frac{\kappa_{xx} + \kappa_{yy}}{2}-\left(\kappa_{xx} - \kappa_{yy}\right) \frac{\delta_x^2-\delta_y^2}{2\delta^2} - \kappa_{xy} \frac{2\delta_x\delta_y}{\delta^2} 
\ee
\be
\kappa_{xy}^\prime = \kappa_{xy} \frac{\delta_x^2-\delta_y^2}{\delta^2} - \left(\kappa_{xx} - \kappa_{yy}\right) \frac{\delta_x\delta_y}{\delta^2} 
\ee
\be
\epsilon_\perp = -\epsilon_x\frac{\delta_y}{\delta} + \epsilon_y\frac{\delta_x}{\delta}  
\ee
where $\delta_i$, $\epsilon_i$, and $\kappa_{ij}$ are the gradients as shown in
\eqs{epsilondef}{kappadef} in the $i$ and $j$ directions of the arbitrary frame.
Therefore, for \eq{vectoranisotdef}:
\be
\Delta_\parallel = \left(\kappa_{xx} - \kappa_{yy}\right) \frac{\delta_x^2-\delta_y^2}{\delta^2} + \kappa_{xy} \frac{4\delta_x\delta_y}{\delta^2} 
\ee
\be
\Delta_\perp = 2\kappa_{xy} \frac{\delta_x^2-\delta_y^2}{\delta^2} - 2\left(\kappa_{xx} - \kappa_{yy}\right) \frac{\delta_x\delta_y}{\delta^2} 
\ee
The angle angle of rotation to the direction of maximum temperature shown in \eq{angledef}, $\theta^\prime$, is also modified in an arbitrary frame.
\be
\theta = \theta^\prime + \theta_\delta,
\ee
where $\theta_\delta$ is the angle between the arbitrary frame and the frame
aligned the temperature gradient (in which we can take advantage of
\eqsc{genanisotdef}{angledef}):
\be
\label{deltaangledef}
\theta_\delta = \text{tan}^{-1}\left( \frac{\delta_y}{\delta_x}\right)
\ee
\section{Anisotropy calculation}
	\label{arbitraryanisot}
Given a 2D temperature tensor in an arbitrary frame, the anisotropy can be calculated as:
\be
A = \frac{2T_{ani}}{T_{iso}-T_{ani} } 
\ee
where,
\be
T_{ani} =\frac{1}{2}\left(\left(T_{xx}-T_{yy}\right)^2 + 4 T_{xy}^2\right)^{1/2}  
\ee
\be
T_{iso} =\frac{T_{xx}+T_{yy}}{2}  
\ee
The angle between an arbitrary frame and the frame indicated by the double prime
notation (see \eq{doubleprime}), where we diagonalize the in-plane components of the the temperature tensor $T_{ij}$,
is:
\be
\theta = \frac{1}{2} \text{tan}^{-1}\left(\frac{T_{xx} - T_{yy}}{T_{xy}}\right)
+ \frac{\pi}{2}\left(1-\Theta\left(T_{xx}-T_{yy}\right)\right)
\ee

Alternatively, the anisotropy can be expressed in vector form to indicate the
direction where the temperature is hottest, without calculating the angle.

\be
A_x = A\left(\frac{
T_{ani} + T_{xx} -T_{yy}}{2 T_{ani} } \right)^{1/2}
\ee
\be
A_y = A{\left(\frac{T_{ani} - T_{xx}+T_{yy}}{2 T_{ani} }\right)}^{1/2} \text{sign}\left(T_{xy}\right) 
\ee
\section{Density and temperature distributions}
	\label{nTdistribution}
The initial density and temperature distributions used in the simulations are the following:
\begin{equation}
\label{distributionfunction}
\begin{array} {l}
n = \begin{cases} (n_0-n_b)
\cos(\pi r/2L_T)^2 + n_b, & \mbox{if } r < L_T, \\
n_b, & \mbox{otherwise}, \end{cases} \\ \\
v_T = \begin{cases} (v_{T0}-v_{T0b}) \cos(\pi |x|/2L_T)^2 + v_{T0b}, & \mbox{if } |x| < L_T, \\
v_{T0b}, & \mbox{otherwise}, \end{cases} \\
\end{array}
\end{equation}
\be
\mbox{where } r = \sqrt{x^2+{(L_T/L_n y)}^2}, \nonumber
\ee
and we take $n_b = 0.1 n_0$
and $v_{T0b} = 0.05 v_{T0}$.
For the Weibel simulation $|x|$ is replaced with $r$ in \Eq{distributionfunction}.
Note that in Ref.~\cite{Schoeffler14} and Ref~\cite{Schoeffler16}, this was erroneously expressed as:
\begin{equation}
\begin{array} {l}
n = \begin{cases} (n_0-n_b)
\cos(\pi r/2L_T) + n_b, & \mbox{if } r < L_T, \\
n_b, & \mbox{otherwise}, \end{cases} \\ \\
v_T = \begin{cases} (v_{T0}-v_{T0b}) \cos(\pi |x|/2L_T) + v_{T0b}, & \mbox{if } |x| < L_T, \\
v_{T0b}, & \mbox{otherwise}, \end{cases} \\
\end{array}
\end{equation}
where the square was omitted.

\end{appendices}

%

\end{document}